\documentclass[aps,prb,twocolumn,superscriptaddress,floatfix]{revtex4-2}

\usepackage{times}
\usepackage{mathptmx}
\usepackage{amsmath}
\usepackage{txfonts}
\usepackage{bm}
\usepackage[dvipdfmx]{graphicx}
\usepackage{bmpsize}
\usepackage{tabularx}
\usepackage{natbib}
\usepackage[dvipdfmx]{hyperref}
\begin{document}

\title{Quantized Conductance by Accelerated Electrons}


\author{D. Terasawa}
\email[]{terasawa@hyo-med.ac.jp}
\affiliation{Department of Physics, Hyogo Medical University, Nishinomiya 663-8501, Japan}


\date{\today}

\begin{abstract}

One-dimensional quantized conductance is derived from the electrons in a homogeneous electric field by calculating the traveling time of the accelerated motion and the number of electrons in the one-dimensional region. 
As a result, the quantized conductance is attributed to the finite time required for ballistic electrons to travel a finite length.
In addition, even if the conductance is finite, it is possible to say that this model requires no Joule heat dissipation,  because the electrical power is converted to kinetic energy of electrons.
Furthermore, the relationship between the non-equilibrium source-drain bias $V_\mathrm{sd}$ and the wavenumber $k$ in a one-dimensional conductor is shown as $k \propto \sqrt{V_\mathrm{sd}}$. 
This correspondence explains the wavelength of the coherent electron flows emitted from a quantum point contact. 
It also explains the anomalous $0.7 \cdot 2e^2/h$ ($e$ is the elementary charge, and $h$ is the Plank's constant) conductance plateau as a consequence of the perturbation gap at the crossing point of the wavenumber-direction-splitting dispersion relation.
We propose that this splitting is caused by the Rashba spin-orbit interaction induced by the potential gradient of the quantum well at quantum point contacts. 
\end{abstract}

\maketitle

\section{Introduction}

Since the successful interpretation of electric conductivity by Drude\cite{Drude1,*Drude2} and Sommerfeld\cite{Sommerfeld}, our common understanding of electric current is based on the macroscopic number of electrons moving at a uniform average velocity.
That is, this model assumes that the electrons are scattered by impurities or phonons.
However, technological progress has enabled us to measure the ballistic, scatterless conduction of electrons. 
In one-dimensional (1D) ballistic conduction, the conductance is quantized in units of $2e^2/h$\cite{vanWees,Wharam}, where $e$ denotes the elementary charge, and $h$ denotes Planck's constant. 
Therefore, we are faced with a finite conductance, despite the absence of scattering and a universal value independent of the sample geometry or quality. 
In order to explain the quantization, the Landauer formula\cite{Landauer,*Landauer_PhilMag,FisherLee,Langreth,ButtikerImryLandauerPinhas,Buttiker,ImryLandauer_Review,Datta} is applied, however, the explanation derived using this model contains controversial issues\cite{Langreth,Datta,Vignale,MukundaDas,Kawabata_Review,Kawabata,Maslov,Micklitz_PRB,Micklitz_PRL,SoreeMagnus,Geng}.
In addition, the existing theories cannot solve the essential question of how and where the Joule heat dissipation of this finite conductance occurs.
Furthermore, we still have a long-term disputed feature in the 1D conduction in the framework of the Landauer formula: the mysterious $0.7\cdot 2e^2/h$ conductance plateau\cite{Micolich}, often called the 0.7 anomaly. 
Considering that the electric current is the electric charges carried per unit time (A = C$\cdot$s$^{-1}$ in the SI unit), we think that the finite conductance arises from the finite time that is necessary to carry charges for a finite length. 
However, it seems that the discussion on this point has not been sufficiently presented thus far. In particular, the time required for ballistic transport has not been seriously considered; ballistic transport is a collisionless transport, thus, if an exerting force is applied to electrons,  these electrons are continuously accelerated.
Recently, an article counting the number of accelerated electrons in a homogeneous electric field\cite{NumOfElectrons} indicates a possible quantization of conductance, however, whether it applies to real systems is an open question.

In this study, we consider the accelerated electrons in a actual situation to derive the 1D quantized conductance. 
First, we discuss the conditions of wavenumbers for the quantized levels in the 1D region.
Then, we numerically show that the electrostatic potential for a two-dimensional (2D) plane with a narrow bridge in the center shows that the potential decreases linearly along this bridge region, showing a homogeneous electric field.
Then, by considering the Schr\"odinger equation in a homogeneous electric field and the Ehrenfast theorem, we derive the time necessary for an accelerated electron to propagate through the 1D region. 
Furthermore, by considering the number of electrons in the 1D region that arrive within this time, we derive the quantized conductance. 
Consequently, we find that the quantized conductance arises from the uniform accelerated motion of electrons.
We also propose how to resolve the Joule heat dissipation in this acceleration model.

Subsequently, we compare this model to an actual system, namely a narrow 1D constriction using a split-gate, known as a quantum point contact (QPC), in a two-dimensional electron system (2DES).
After discussing how this model can represent the conduction of QPCs, we show that the wavelength of coherent electron flows emitted by a QPC\cite{Topinka_Science,Topinka_Nature,Braem,Brun,Kozikov} corresponds to the de-Broglie wavelength derived from this model.
Furthermore, we show that this model explains the anomalous $0.7\cdot 2e^2/h$ plateau. 
Using the relationship between the non-equilibrium voltage and the wavenumber that is derived in this model, we show that the quadratic dispersion relation between the energy and the wavenumber changes to the linear relation between the energy and the voltage. 
Then, we successfully demonstrate the peculiar subband edge (SBE) lines that are concurrent with the $0.7\cdot 2e^2/h$ conductance plateau.
As a result, we explicitly show that the 0.7 anomaly originates in the perturbation gap of the bands, which split in a wavenumber direction due to the Rashba-type spin-orbit interaction.

\section{Quantized conductance by accelerated electrons}
\label{Sec_AccelerateModel}

In this section, we propose a new model that derives the quantized conductance by assuming a uniform accelerated motion of electrons. Before we unfold our model, a brief introduction to the Landauer formula is in order, and discuss a previously unaddressed issue in Landauer's model.
We also simulate the initial electrostatic potential for a two-dimensional plane with a narrow constriction by solving the Poisson equation.

\subsection{Brief Introduction of Landauer's model}

\begin{figure} 
\includegraphics[width=0.8\linewidth]{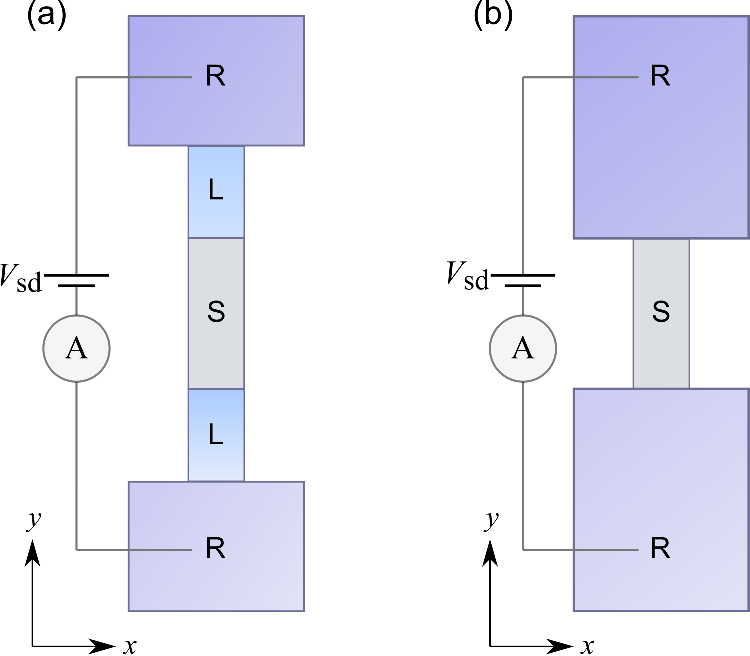}
\caption{\label{fig_model}(a) Landauer's model and (b) the model considered here. R, L, and S stand for the reservoir, lead, and sample, respectively.}
\end{figure} 

In 1957, Landauer cast doubts on understanding conductance in terms of the equation of motion\cite{Landauer}, because uniform distribution in the electric field is merely expected due to inhomogeneity in the conductor.
Instead, Landauer aspired to understand conductance in terms of transmission and reflection at the boundaries\cite{Landauer,Landauer_PhilMag,ImryLandauer_Review}.
In Landauer's model, a sample conductor is sandwiched between two leads, each of which is connected to a reservoir, and the source-drain bias voltage $V_\mathrm{sd}$ is applied across the reservoirs, as shown in Fig.\,\ref{fig_model} (a).
In this model, non-interacting electrons with wavenumbers corresponding to the energy range between the source and drain enter the sample region owing to the non-equilibrium bias.  
Denoting the 1D energy dispersion as $E(k)$, the current $i_k$ carried by the electrons at wavenumber $k$ becomes
\begin{equation}
i_k = \frac{e}{L_N}\frac{dE(k)}{\hbar dk},
\end{equation}
where $L_N$ describes the normalization length of the 1D wavefunction, and $dE(k)/\hbar dk$ represents the group velocity.
The total current $I$ is carried by the electrons in the state between the source and drain chemical potentials, $\mu_\mathrm{s}$ and $\mu_\mathrm{d}$, respectively, thus
\begin{equation}
I = \int_{k_\mathrm{d}}^{k_\mathrm{s}} i_k L_N \frac{dk}{2\pi} = \frac{e}{2\pi\hbar}\int_{k_\mathrm{d}}^{k_\mathrm{s}}\frac{dE(k)}{dk}dk = \frac{e}{h}\int_{\mu_\mathrm{d}}^{\mu_\mathrm{s}}dE = \frac{e}{h}(\mu_\mathrm{s} -\mu_\mathrm{d}),   \label{LandauerI}
\end{equation}
where $k_\mathrm{s}$ and $k_\mathrm{d}$ represent the wavenumbers of the source and drain, respectively. Considering that $\mu_\mathrm{s} - \mu_\mathrm{d} = eV_\mathrm{sd}$, the conductance $G$ becomes\cite{ImryLandauer_Review,Kawabata_Review}
\begin{equation}
G=  \frac{e^2}{h}. \label{LandauerFormula}
\end{equation}
If we consider the transmission between the lead and the sample, the conductance can be written as
\begin{equation}
G= \frac{e^2}{h} \mathcal{T},
\end{equation}
where $\mathcal{T}$ denotes the transmission coefficient.
Experimentally, the quantized conductance is observed in units of $2e^2/h$\cite{vanWees,Wharam,Frank}, in which it is recognized that $\mathcal{T}$ reaches an ideal case ($\mathcal{T}=1$) with degenerated spin degrees of freedom.
Later, this formula was expanded to a multi-channel case to understand mesoscopic conductance fluctuations\cite{Tankei,Higurashi,Tamura}.

In Landauer's model, the finite conductance is attributed to inevitable contact resistances at the interfaces of the reservoirs and leads\cite{ImryLandauer_Review,Datta}, and the Joule heat dissipation due to the finite conductance is interpreted to occur inside the reservoirs (not at the interfaces of the reservoirs and leads)\cite{ImryLandauer_Review,Datta,Rokni}.
However, this model artificially assumes that a voltage drop occurs at the interfaces between the sample and leads\cite{Datta}, and equilibration occurs inside the sample.
%
Furthermore, a similar formulation can be derived using the Kubo formula even when no reservoirs are assumed\cite{Economou,Thouless,Kawabata_Review}.
Thus, it is controversial to interpret the quantized conductance (in the limit of $\mathcal{T} = 1$ in Eq.\,(\ref{LandauerFormula})) as the contact resistance.
Several other studies also questioned about Landauer's model\cite{Micklitz_PRB,Micklitz_PRL,MukundaDas,SoreeMagnus,Geng}, and some of them deduced the quantized conductance using different methods\cite{MukundaDas,SoreeMagnus,Geng}.
However, these studies\cite{MukundaDas,SoreeMagnus,Geng}, which assume a constant speed in the QPC region, are inevitably attached by an attenuation term similar to the transmission coefficient of the Landauer formula; thus, they need to be scrutinized as to whether the term satisfies unity. 
However, we no longer examine the validity of conventional interpretations and assumptions.

\subsection{Initial Conditions for 1D Conduction}

\begin{figure} 
\includegraphics[width=1\linewidth]{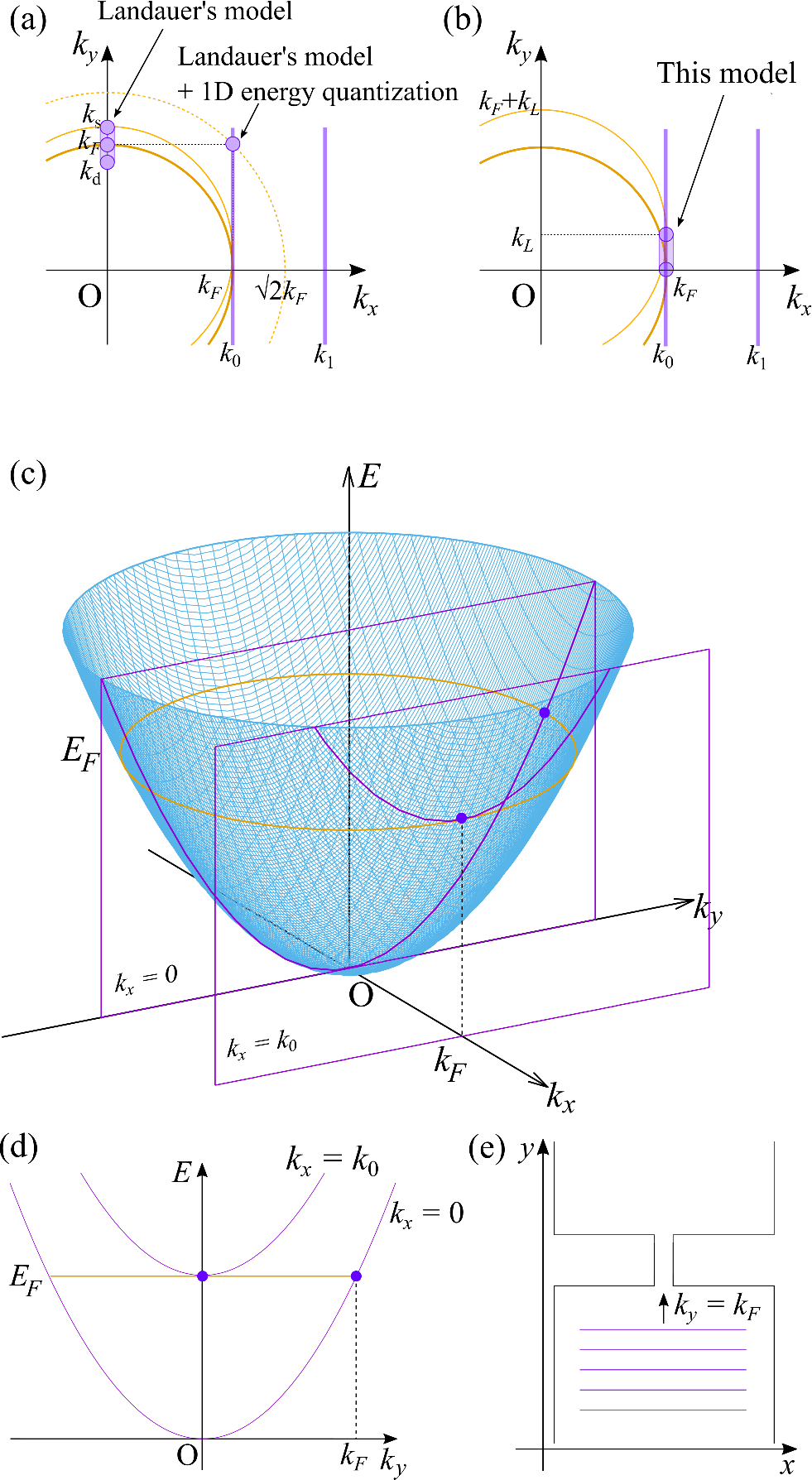}
\caption{\label{fig_FermiCircle}Range of wavenumbers that are considered in (a) Landauer's model and (b) the accelerated model (only for $k_x \geq 0$) when the Fermi surface reaches the zeroth energy of the quantum harmonic oscillator. $k_0$ and $k_1$ represent the wavenumbers that are equivalent to the zeroth and first energies of the quantum harmonic oscillator. (c) 2D dispersion relation with line cuts at $k_x =0$ (for Landauer's model) and $k_x =k_0$ (for this model) planes, when $E_F$ is equal to the zeroth harmonic oscillator energy $E_0$ ($k_0 = k_F$). (d) 1D $y$-directional electron dispersion relations for $k_x=0$ and $k_x =k_0 (=k_F)$ cases. (e) An electron plane wave with $\bm{k}=(0,k_F)$ in a 2D system.}
\end{figure}

Herein, we would like to consider what the initial conditions for 1D conduction are.
As shown above, Landauer's model assumes that the electrons in the states between $\mu_\mathrm{s}$ and $\mu_\mathrm{d}$ carry the current. 
The Fermi energy, $E_F$, lies between these levels; therefore, this model assumes that the electrons near the Fermi wavenumber, $k_F$, transmit the 1D region. 
However, we need to consider quantization lateral to the propagation direction simultaneously; conductance quantization occurs when the width of the constriction is comparable with the Fermi wavelength $\lambda_F = 2\pi/k_F$.
For simplicity, we assume that the reservoirs and leads are two-dimensional (2D). 
If we consider the $y$-direction as the direction of electron propagation, and the $x$-direction as the direction perpendicular to the propagation, the range of wavenumbers for Landauer's model in the wavenumber vector space can be depicted in Fig.\,\ref{fig_FermiCircle} (a).
As the potential of the $x$-direction in the 1D sample region can be described by the harmonic oscillator, $U(x) = m\omega_x^2x^2/2 $, where $\omega_x$ represents the oscillation frequency, and $m$ denotes the electron mass, 
2D electrons that reach one of the discrete levels, 
\begin{equation}
E_n = \left( n+\frac{1}{2}\right) \hbar\omega_x, \ \ n=0,1,2,\cdots, 
\end{equation}
can transmit through the 1D region. 
The semi-classical wavenumber $k_n$ with energy equal to the energy of $n$-th quantum state ({\it not}  quantized wavenumber\cite{note_quantizedwavenumber}) can be described as 
\begin{equation}
k_n = \sqrt{ (2n+1)\frac{m\omega_x}{\hbar}}, \ \ n=0,1,2,\cdots.
\end{equation}
The first two wavenumbers $k_0$ and $k_1$ are shown in Fig.\,\ref{fig_FermiCircle} (a) and (b).
Since the quantization condition is in the $x$-direction, the first electron in the 2DES that reaches this condition is the electron with a wavenumber vector of $\bm{k} = (|k_F|, 0)$.
Remember that the wavefunction is basically the superposition between $k_x = k_\mathrm{F}$ and $-k_\mathrm{F}$ in the confinement (localization) state, although Fig.\,\ref{fig_FermiCircle} (a) to (c) show only the $k_x > 0$ state.
Consequently, considering that $k_y$ is also $\simeq k_F$ in Landauer's model, the wavenumber vector becomes $\bm{k} = (k_F, k_F)$, which ends up exceeding the magnitude of the Fermi wavenumber by approximately $\sqrt{2}$ times ($|\bm{k}| =\sqrt{2}k_F$), as shown in Fig.\,\ref{fig_FermiCircle} (a). 
This corresponds to twice the Fermi energy $E_F$, which cannot be possible.
Of course, the electron must be in the Fermi level $E_F$; therefore, to satisfy $|\bm{k}| = k_F$ when it enters the 1D conduction region, we must require $k_y =0$.
In other words, electrons with $\bm{k} = (0,k_F)$ in the 2D region (electrons in the long-wavelength limit perpendicular to the propagation direction, i.e., a plane wave) cannot enter the 1D region as is (see Fig.\,\ref{fig_FermiCircle} (e)).

\begin{figure} 
\includegraphics[width=1\linewidth]{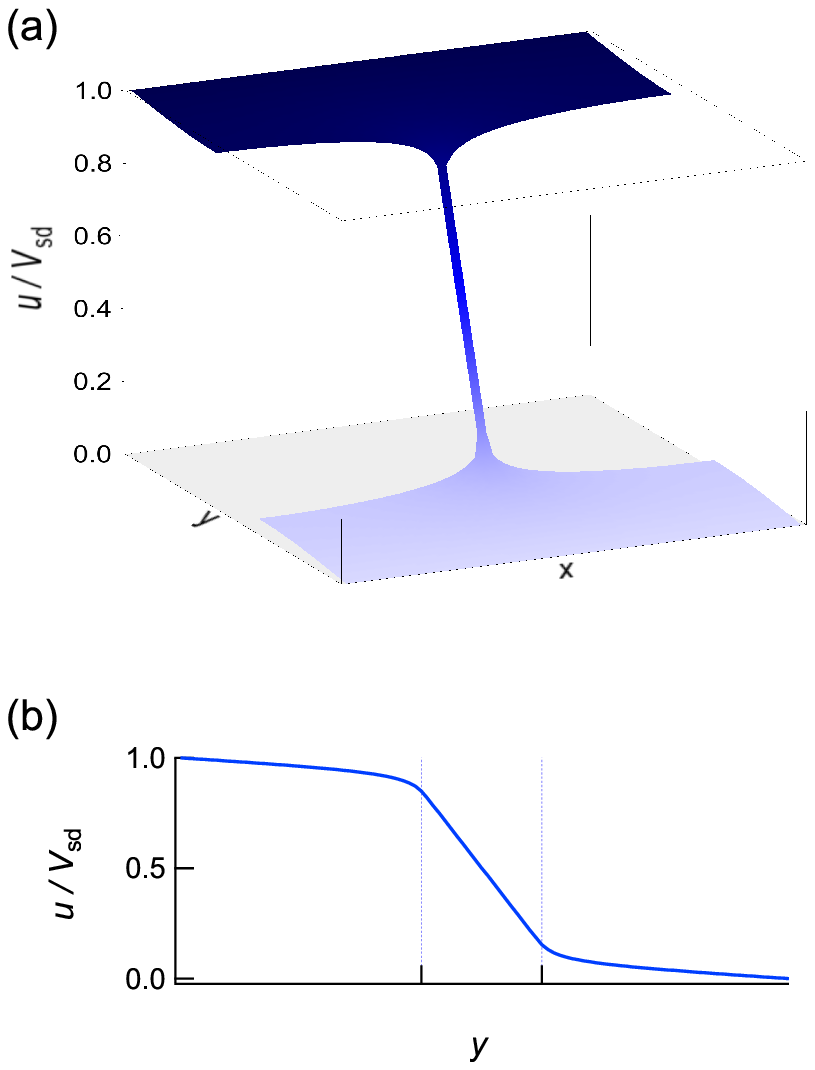}
\caption{\label{fig_Poisson}\,(a) Simulation result of the electric potential for a 2D plane with a narrow constriction region as a solution of the Poisson equation, and (b) the line profile of the potential change at the center line.
}
\end{figure}

To further understand the initial condition, we subsequently compute the 2D Poisson equation (Laplace equation as the right-hand side is zero)
\begin{equation}
\nabla^2u(x,y)=0
\end{equation}
where $u(x,y)$ is the electric potential, to simulate the 2D electric potential with a narrow constriction. 
First, we prepare a 2D plane with a narrow bridge region in the center of the plane.
Then, a uniform electric potential is prepared ($u/V_\mathrm{sd}=0.5$), with one end connected to ground ($u(x,N_\mathrm{grid}) =0$, where $N_\mathrm{grid}$ is the number of the grids in one direction) and the other end connected to a constant positive voltage ($u(x,0)/V_\mathrm{sd}=1$) to ground.
Then, we solve the potential change self-consistently using the finite difference method. 
Figure \ref{fig_Poisson}\, (a) shows the result of the calculation and (b) shows the line profile at the center point along the $y$ axis (see gif animations in the supplementary material).
The result differs from the assumed potential change of Landauer's model;
in a standard interpretation of Landauer's model, the potential is assumed to be flat in the 1D region and to drop at both ends of the 1D region\cite{Datta}, however, the result shows a linear potential decrease in the narrow constriction region.
This result is similar to the result of a recent study of graphene using the scanning tunneling potentiometry\cite{ZJKrebs}, which shows a linear potential decrease in the narrow conduction region.
Therefore, we need a new model that takes this potential change and the above initial condition into account.

\subsection{Accelerated electrons model}

\begin{figure} 
\includegraphics[width=1\linewidth]{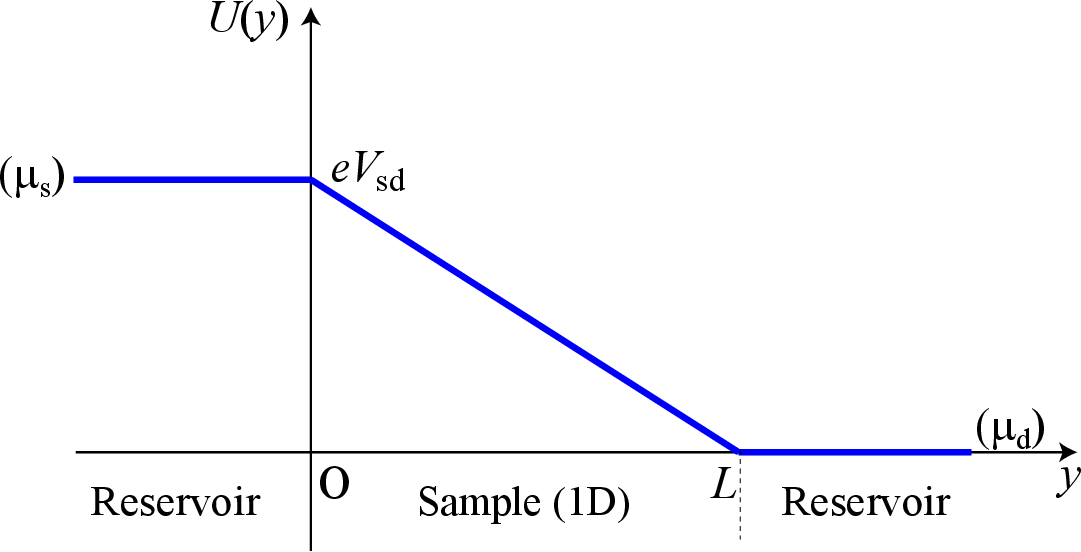}
\caption{\label{fig_potential}The potential of our model (a homogeneous electric field potential in the sample region).}
\end{figure}

We consider a 1D conductor with a length of $L$, and the source and drain reservoir electrodes to be attached directly to the 1D conductor.
Then, a small DC bias $V_\mathrm{sd}$ is applied across the electrodes (see Fig.\,\ref{fig_model} (b)). 
Then, as shown in the simulation result, we expect that the this bias make a homogeneous electric field $V_\mathrm{sd}/L$ in the 1D conductor. 
For simplicity, we describe the potential $U(y)$ as 
\begin{equation}
U(y) = 
\begin{cases}
eV_\mathrm{sd}  &  y < 0 \\
-\frac{eV_\mathrm{sd}}{L}y +eV_\mathrm{sd} &  0 \leq y \leq L \\
0  & y > L 
\end{cases}
\end{equation} 
as shown in Fig.\,\ref{fig_potential}.
Then, we consider an electron that travels through the conductor. 
As an essential assumption, electrons can travel ballistically, and the contacts between the reservoirs and the 1D conductor are ideal; thus, no scattering or reflection is expected during conduction.
As discussed in the previous subsection, the electron is considered to be stationary ($k_y =0$) at the starting point immediately after it is emitted from the source.
The electron is subsequently accelerated by a constant electric field, such as electrons in an electrostatic accelerator. 
In this case, if the electron needs time $\tau$ to traverse the conductor from the source to the drain, the electric current carried by an electron $j$ can be written as 
\begin{equation}
j = \frac{e}{\tau}. \label{j=e/tau}
\end{equation}
To deduce this time $\tau$ in the non-relativistic regime, we consider the electron motion in a homogeneous electric field.

In quantum mechanics, the coordinate space Hamiltonian in the 1D region becomes 
\begin{equation}
\hat{H}_y= \frac{\hat{p_y}^2}{2m} -\frac{eV_\mathrm{sd}}{L}y +eV_\mathrm{sd},
\end{equation}
where $\hat{p_y}$ is the momentum operator.
This model is attributed to solving the time-independent Schr\"odinger equation in a linear potentail, which reduces to the Airy equation in the coordinate space. 
This gives rise to an exotic stationary solution, which is known as the Airy function\cite{LandauLifshitzVol3,Griffiths}, unlike the semi-classical solution. 
However, it is known that a time-dependent solution can be derived from the Fourier transform of the momentum space solution\cite{Nauenberg}.
In the momentum space, the position operator becomes $\hat{y} = i\hbar\frac{\partial}{\partial p_y}$ and the Hamiltonian in the momentum representation is described as 
\begin{equation}
\hat{H}_{p_y}  = \frac{p_y^2}{2m} -i\hbar F\frac{\partial}{\partial p_y} +E_0,
\end{equation}
where $E_0=eV_\mathrm{sd}$, and $F = -\nabla U(y) = eV_\mathrm{sd}/L$. 
The Schr\"odinger equation for this problem is
\begin{equation}
-i\hbar F \frac{\partial \phi}{\partial p_y} + \left(\frac{p_y^2}{2m} + E_0-E\right)\phi =0, \label{SchrEq_p}
\end{equation}
where $E$ denotes the energy, and $\phi$ denotes the wavefunction in the momentum representation.
Because the system is not in a bound state, the motion does not change after considering the general solution in an infinitely long homogeneous electric field and then applying the initial energy conditions.
Eq.\,(\ref{SchrEq_p}) is a first-order partial differential equation and its general solution becomes\cite{LandauLifshitzVol3,Nauenberg,Robinett}
\begin{equation}
\phi(p_y,E) = C \exp \left[ \frac{i}{\hbar F} \left(-(E_0- E )p_y -\frac{p_y^3}{6m}\right) \right],     \label{a_E}
\end{equation}
where $C$ is a constant. 
Then, the Fourier transform of Eq.\,(\ref{a_E}) is\,\cite{Nauenberg}
\begin{equation}
\psi(y,t) = C \int_{-\infty}^{+\infty}\frac{dp_y}{2\pi \hbar} \int_{-\infty}^{+\infty}\frac{dE}{2\pi \hbar} \phi e^{i(p_yy -E t)/\hbar}.
\end{equation}
The integration over the variable $E$ yields $2\pi\delta((p_y-tF)/F\hbar)$, thus we obtain 
\begin{equation}
\psi(y,t) = \frac{CF}{2\pi \hbar}\exp \left[ \frac{iFt}{\hbar} \left(y - \frac{Ft^2}{6m}-\frac{E_0}{F} \right) \right].       \label{WaveFunction}
\end{equation}
We can readily confirm that this wavefunction is an eigenfunction of the momentum operator
\begin{equation}
 \hat{p_y} \psi = -i\hbar \frac{\partial}{\partial y}\psi = Ft \, \psi.
\end{equation}
For the momentum representation wavefunction, the eigenvalue of the position operator is  
\begin{equation}
 \hat{y} \phi = i\hbar \frac{\partial}{\partial p_y} \phi = \left( \frac{E_0 - E}{F} + \frac{p_y^2}{2mF} \right)\phi.
\end{equation}
In this model, the initial condition is $E=E_0(=eV_\mathrm{sd})$ at $y=0$, thus we obtain the eigenvalue of $p_y^2/(2mF)$.
This form satisfies the conditions of continuity of $\psi$ and $\partial \psi/\partial y$ at $y=L$, for $y >L$ the wavefunction obviously has the form of 
\begin{equation}
\psi = \frac{CF}{2\pi\hbar}\exp \left[ \frac{i}{\hbar}\sqrt{2mE_0}y + \delta \right],
\end{equation}
where $\delta$ is a constant. The traveling time $\tau$ can be derived by putting $y=L$. 

Otherwise, the expectation values of the position and momentum are derived from the Ehrenfest's theorem, 
\begin{equation}
m\frac{d^2\langle y \rangle}{dt^2} = \frac{d\langle p_y \rangle}{dt} = F.
\end{equation}
Hence, we obtain
\begin{gather}
\langle y \rangle =\frac{Ft^2}{2m},        \label{x-t} \\
\langle p_y \rangle = Ft,                        \label{p-t}
\end{gather}
when $E=E_0$ and $p_y = 0$ at $t=0$.
As a result, we obtain eigenvalues and expectation values that are the same in form as the classical Newton's equation of motion.
From the relationship between time and position, we obtain the traveling time:
\begin{equation}
\tau = \sqrt{\frac{2m}{E_0}}L = \sqrt{\frac{2m}{eV_\mathrm{sd}}}L.                   \label{tau}
\end{equation}
Therefore, Eq.\,(\ref{j=e/tau}) becomes
\begin{equation}
j =  \frac{e}{L} \sqrt{\frac{eV_\mathrm{sd}}{2m}}.
\end{equation}


Subsequently, we derive the number of electrons that can exist in a 1D conductor.
The electrons are assumed to be constantly supplied by the source electrode and retrieved by the drain electrode via $V_\mathrm{sd}$.
Each electron performs the same motion, and the same number of electrons reaches the drain for every $\tau$ of elapsed time.
This motion is not constrained, thus the electrons have continuous momentum eigenvalues corresponding to the non-degenerated continuous energy spectrum occupying from 0 to $eV_\mathrm{sd}$. 
Therefore, the number of electrons $N$ can be determined by maximizing the occupancy of electrons in this energy interval. 
After all, when the spin is degenerated, this can be determined by the one-dimensional density of states as follows:
\begin{equation}
N = 2\int_0^{k_L} L \frac{dk_y}{2\pi} =\frac{2L\sqrt{2meV_\mathrm{sd}}}{2\pi\hbar}, \label{NumberofElectrons}
\end{equation}
where $k_L$ denotes the wavenumber immediately before the electron reaches the drain, which is derived using the momentum expectation value from Eq.\,(\ref{p-t}) as
\begin{equation}
k_L = \frac{p_L}{\hbar} =\frac{\sqrt{2meV_\mathrm{sd}}}{\hbar}.          \label{wavenumber}
\end{equation}
We obtain the same result by considering the integration of the normalized coefficient derived from the delta-function normalization\cite{NumOfElectrons}.
Consequently, from the definition of electric current, the total amount of current $J$ carried by these electrons becomes
\begin{equation}
J = \frac{eN}{\tau} = \frac{2e^2}{h}V_\mathrm{sd}.          \label{Current}
\end{equation}
Therefore, we obtain 
\begin{equation}
G = \frac{2e^2}{h}.       \label{ConductanceQuantum}
\end{equation}
The above calculation assumes an initial velocity of zero; however, the quantization holds even if the initial velocity has a finite value.
Furthermore, by adapting the same argument, the quantized conductance is also derived in the case of a massless linear band (see Appendix \ref{LinearBand}).

As mentioned earlier, for an electron to transmit the 1D region, the Fermi energy must reach a quantized level in the $x$-direction.
Considering this condition, the wavenumber in the $x$-direction is comparable to $k_F$ when entering the 1D region. 
Thus we require that $k_y =0$ at the starting point, since the system must satisfy $|\bm{k}| = k_F$ with $k_x \simeq |k_F|$.
Then, $k_y$ increases to $k_y=k_L$ with $k_x$ constant ($ \simeq |k_F|$).
Figure \ref{fig_FermiCircle} (b) shows this condition in wavenumber space, and (c) and (d) show the dispersion relation for 1D conduction. 
The Fermi statistics also requires $k_y$ to be zero, because the $0 < k_y \leq k_L$ region is occupied by the previously departing electrons. 
However, none of the electrons less than $|k_x| < k_F$ can contribute to the electric conduction (if the argument is restricted to the lowest energy level), because these electrons cannot enter the 1D region.
In the present discussion, the motion of the wavepackets has not been considered.
However, it has been shown that the center of the wavepackets moves according to the accelerated motion\cite{Nauenberg,Robinett}, therefore, we believe that the results do not change if the wavepacket is considered.

Consequently, the quantized conductance is derived from the equation of motion and, unlike the Landauer formula, the conductance is intrinsic to the ballistic 1D conductor; it does not originate in the contact resistances.
Instead, the finite conductance arises from the finite time required for electrons to travel a finite distance.
Notice the difference between the superconductor and this model. In contrast to this model, there is no voltage drop in the superconductor. 
In addition, the difference between our model and the Drude model becomes clear.
The Drude model, as well as the Landauer model, considers a macroscopic number of electrons moving at a constant averaged velocity.
However, our model considers the linear motion of uniform acceleration: electrons free-fall over a distance of only $\approx 100$\,nm for a QPC in a 2DES.
The concept of a constant averaged velocity is adopted in the Kubo formula\cite{FisherLee,BarangerStone,Kawabata_Review} and several other studies\cite{Geng}, which basically treat the non-equilibrium effect as a perturbation method; thus, they consider the motion of constant velocity in the equilibrium limit, and then derive the conductance as a linear response to a time-dependent external field. 
However, we assume that the field is time-independent and so is the Hamiltonian, and thus the system is in a quasi-equilibrium state with no energy dissipation.
We will discuss the timescale which switches between the conduction region and non-conduction region in Sec.\,\ref{QPCsIn2DES}.

\subsection{Joule heat}
\label{JouleHeat}

In Landauer's model, the question of how and where the Joule heat dissipation occurs is another controversial issue.
In a standard interpretation, it is assumed that the heat dissipation takes place in the reservoirs.
Das and Green\cite{MDasGreen} speculated that the dissipative process is caused by many-body electron-hole effects; however, they did not show any practical formulations or evidences.
In this study, the authors wrote, ``Any finite conductance $G$ must dissipate electrical energy at the rate $P=IV=GV^2$, where $I=GV$ is the current and $V$ the potential difference across the terminals of the driven conductor.'' In addition to this statement, they  wrote, as the requirement of thermodynamic stability, ``ballistic quantum point contacts have finite $G \propto e^2/\pi\hbar$; therefore, the physics of energy loss is indispensable to any theory of ballistic transport''. 
As can be seen from this description, previous theories based on the Landauer formula thus far have been limited to explaining the Joule heat dissipation in the absence of a scattering mechanism.

Here, we would like to propose a possible interpretation of the Joule heat dissipation using this model.
Importantly, as shown in the previous subsection, this model shows that the finite conductance (finite resistance) does NOT originate in the scattering events, and that the conducting electrons are all accelerated by the electric force.
These points suggest that a finite conductance does not necessarily involve the energy loss in the form of Joule heat dissipation, if instead the applied electric force is used only to increase the kinetic energy of the electrons.
Note that Ohm's law still holds for the accelerated model, because the resistance in Ohm's law is ultimately just a ratio of current and voltage values and does not refer to any physical reality.
Therefore, we can consider the Joule heat dissipation $Q$ for two distinctive conduction cases:
\begin{equation}
Q = 
\begin{cases}
	0 & \text{for accelerated model} \\
	J^2/G & \text{for Drude model} 
\end{cases}
\end{equation}
Similarly, we should reconsider the meaning of the electrical power $P=J^2/G$. Again, if we consider the unit of this quantity, we obtain J$\cdot$s$^{-1}$ (SI units). 
This only indicates the work rate done by the electric field; it does not indicate what form of energy this power is converted to. 
In the case of accelerated conduction, $P$ is used to increase the kinetic energy of the electrons $K$, while in the case of Drude conduction, it is consumed by the Joule heat dissipation $Q$:
\begin{equation}
P  \longrightarrow 
\begin{cases}
	K & \text{for accelerated model} \\
	Q & \text{for Drude model} 
\end{cases}
\end{equation}
Therefore, the quantized conductance is unlikely to cause any Joule heat dissipation (adiabatic process); the heat is dissipated elsewhere, where the usual Drude model dominates the electrical conductivity.

This was well exemplified in a carbon nanotube resistor that was able to withstand a current density of more than $10^7$\,A$\cdot$cm$^{-2}$\cite{Frank}. 
The authors estimated that the extreme heat reached in the nanotube would be 20,000\,K, which is certainly not possible, and speculated that the heat must be dissipated elsewhere.
In our model, we can conclude that the Joule heat dissipation did not occur in the 1D nanotube conductor (accelerated conduction region), but occurred in the attached metal region (Drude model region).
  
In recent years, Joule heat dissipation has been studied from the viewpoint of hydrodynamic electron flow\,\cite{Tikhonov,Asafov}, and we see many interesting conclusions. 
Therefore, we must discuss this important topic further. Here, we limit ourselves to discussing logical possibilities, and further rigorous theoretical analysis should be done. 

\subsection{QPCs in two-dimensional systems}
\label{QPCsIn2DES}

In a realistic situation, quasi 1D systems such as QPCs are implemented on a high-mobility 2DES. 
At $V_\mathrm{sd}=0$, the electrons in the 2DES regions move in random directions at the Fermi velocity.
When $V_\mathrm{sd}$ is applied, the electrons in the 2DES immediately move in the direction of the applied electric field, and if $E_0 (=\frac{1}{2}\hbar\omega_x) > E_F$ ($G=0$ case), the electrons on the source side encounter the potential barrier under the gate electrodes.
In contrast, the electrons on the drain side are swept instantly from the drain. 
This case is similar to a capacitor charged by the applied voltage, and a homogeneous electric field is formed in between the two electrodes if there is no depletion field.
The source side is filled with electrons up to the chemical potential of the source, whereas the drain side is empty down to that of the drain, and eventually the potential differs $V_\mathrm{sd}$.
Then, as the 1D channel opens, the lateral quantization at the QPC allows electrons that reach the quantized levels to transmit through the QPC.
Perhaps this situation is similar to an hourglass: the entrance of a QPC is congested by electrons and then the electrons enter the QPC one by one as soon as possible.

In this model, our proposed assumption is effectively satisfied, 
because the number of electrons passing through a QPC is always small compared to the number of electrons in $k_x < |k_F|$ at the source side.
Therefore, the voltage (the chemical potential) on each side is maintained at a constant level during the process, because the renormalized chemical potential is negligible. 
In such a case, the 2DES regions attached to the QPC are considered as reservoirs.
This further supports the experimental fact that the sample geometry and quality can be negligible (if the QPC length is smaller than the mean free path) in the conductance value; wherever a QPC is located, only that part affects the conductance. 
For this to occur, the conductance in the 2DES regions, $G_\mathrm{2DES}$, should be much larger than the conductance in the QPC region, $G_\mathrm{QPC}$, as $G_\mathrm{QPC} \ll G_\mathrm{2DES}$. 
Therefore, good quantization is guaranteed for small numbers of quantized channels.
It should be noted that the assumption of a homogeneous electric field locally violates the charge neutrality condition guaranteed by Landau's Fermi liquid theory, as the typical Thomas-Fermi screening length of the 2DES in GaAs is approximately 5\,nm\cite{Datta}. 
Hence, we may need to discuss electron-electron interactions on the source side of the 2DES; however, the quantized conductance is obtained in the free electron model. 
The possibility of electron-electron interactions will be discussed at the end of Sec.\,\ref{Sec_0.7anomaly}.

\begin{figure} 
\includegraphics[width=0.8\linewidth]{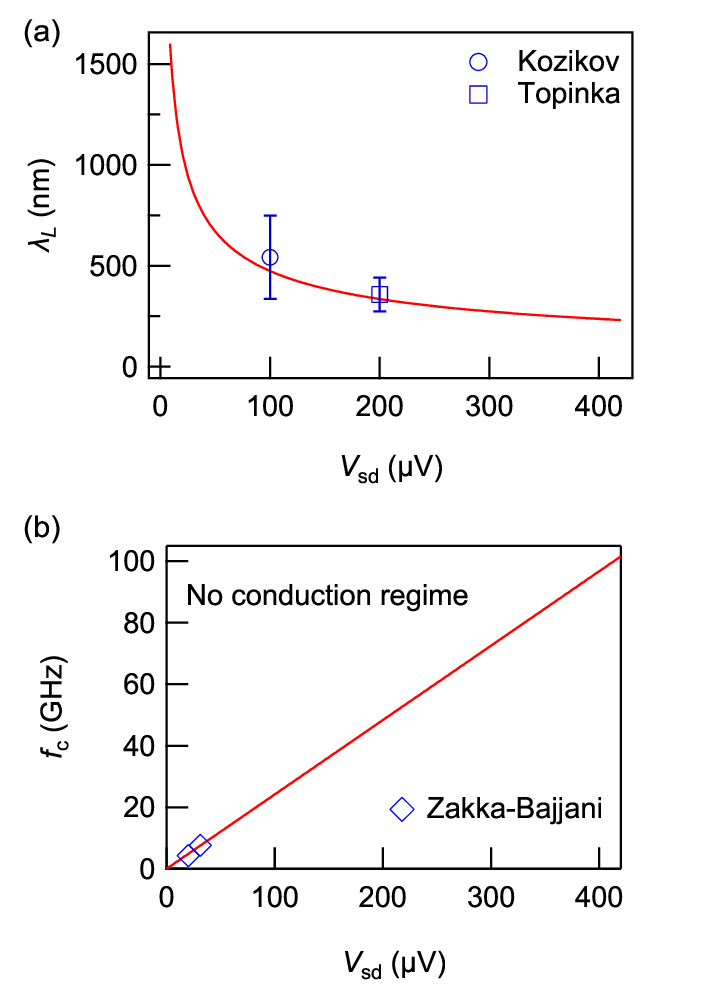}
\caption{\label{fig_lambdaL}\,(a) The de-Broglie wavelength of electrons flowing from a QPC, $\lambda_L$, and (b) the characteristic frequency that limit the conductance quantization, $f_\mathrm{c}$, as a function of $V_\mathrm{sd}$. The conductance quantization is shown in the shaded region. 
In (a), we plot the averaged length of node spacings in electron flows from two SGM experimental results of Kozikov {\it et al}.\cite{Kozikov} and Topinka {\it et al}.\cite{Topinka_Nature} (the data were visually read and collected by the author). 
The averaged value for Ref.\cite{Kozikov} is slightly larger than the value that is mentioned in the main text.
There are several other SGM experiments, however, no comparison could be made because there were no description of $V_\mathrm{sd}$. 
In (b), we plot the threshold $V_\mathrm{sd}$ values for two frequencies in Fig.\,2 of the shot noise results of Zakka-Bajjani {\it et al}.\cite{Zakka-Bajjani}.
}
\end{figure}

Experiments using a scanning gate microscopy (SGM)\cite{Topinka_Science,Topinka_Nature,Braem,Brun,Kozikov} may help to further understand the physics of the proposed model.
These SGM measurements showed coherent electron streams from a QPC.
These electron flows showed node spacings larger than the Fermi wavelength, and significantly smaller than the mean free path, thermal length, and dephasing length. 
In this model, the electrons flooding out of a QPC subsequently flow at a constant velocity until they collide with obstacles, such as impurity potentials or vacancies, or lose their coherence due to, for example, electron-electron scattering. 
The flows are robust, as observed, because the initial conditions of the electrons are expected to be identical.
Furthermore, our model predicts that the electrons out of a QPC possess a de-Broglie wavelength of 
\begin{equation}
\lambda_L = \frac{2\pi}{k_L} = \frac{h}{\sqrt{2m^\ast e V_\mathrm{sd}}}, 
\end{equation}
where $m^\ast$ is the effective mass of the electrons.
Then, there should be another scale of oscillation in addition to the half Fermi wavelength that was observed everywhere in the 2DES.
The observed node spacings are approximately 230\,nm (see the flow branch of the line profile in Fig. 2 of Ref.\cite{Kozikov}), which probably matches this wavelength, $\lambda_L$.
In their experiment\cite{Kozikov}, a source-drain bias of 100\,$\mu$V AC r.m.s. was used to drive the current.
This gave $\lambda_L \approx 470$\,nm, which was approximately twice the value. 
Other than this wavelength, many what appear to be wave oscillations and fringes of approximately 200\,-\,500\,nm were observed in \cite{Kozikov}.
In Fig.\,\ref{fig_lambdaL} (a), we compare $\lambda_L$ with the averaged wavelength obtained from two SGM experimental results. 
In the absence of other promising length scales to explain the observed node spacings (Fermi wavelength $= 72$\,nm, elastic mean free path $=49$\,$\mu$m, thermal length $=9$\,$\mu$m, and dephasing length $=200$\,$\mu$m in \cite{Kozikov}), the wavelength derived from this model deserves consideration.
In addition to this propagating-directional wavelength, attention must be paid to the lateral-directional wavelength, i.e., the width of the electron flow.
It seems obvious that the width corresponds to the width of the QPC, which is approximately the same length as the Fermi wavelength shown in the article. 
This result indicates that the $x$-directional wavenumber is approximately the same as the Fermi wavenumber, which further supports the prerequisite for the accelerated model ($k_x = k_F, k_y = 0$).
Consequently, we can confirm that the total kinetic energy of the electrons in these streams  corresponds to the sum of $E_F$ and $eV_\mathrm{sd}$.

In this experiment\cite{Kozikov}, an AC electric field was applied to the sample, however, the frequency was below several hundreds of Hz (considering the description ``using a standard lock-in technique"). 
For the electrons in a high-mobility sample, the AC field in this frequency range is equivalent to a slowly varying DC field; they equilibrate at every moment.
Thus, the observed quantized conductance should be interpreted as a static DC conductance.
According to this accelerated model, there would be a characteristic frequency 
\begin{equation}
f_\mathrm{c} = \frac{N}{\tau} = \frac{eV_\mathrm{sd}}{h},
\end{equation}
which is also derived from the fluctuation-dissipation theorem\cite{Blanter}, and has been already confirmed by the shot noise experiment using QPC\cite{Zakka-Bajjani}.
Fig.\,\ref{fig_lambdaL} (b) shows this frequency, $f_\mathrm{c}$, as a function of $V_\mathrm{sd}$.
The frequency becomes 24\,GHz for a typical source-drain voltage of 100\,$\mu$V, and the regime in $f > f_\mathrm{c}$ is no conduction regime.
In the no conduction regime, the system can be described by the solution of the Schr\"odinger equation for a time-dependent linear potential\cite{Guedes}, where the position of the electrons oscillates around a point.
Thus far, a high-frequency conductance measurement using a frequency of 160\,MHz ($< f_\mathrm{c}$) has shown that the imaginary part becomes capacitive; however, the real part remains the same as the DC measurement\cite{Regul}.
In the next section, we further show that this accelerated model explains another long-disputed phenomenon regarding the QPC.

\section{0.7 anomaly}
\label{Sec_0.7anomaly}

In the previous section, we have revealed an important relationship between the wavenumber of an electron $k$ and the applied source-drain voltage $V_\mathrm{sd}$ as shown in Eq.\,(\ref{wavenumber}): $k$ is proportional to $\sqrt{V_\mathrm{sd}}$.
As a direct consequence of Eq.\,(\ref{wavenumber}), a quadratic dispersion relation $\propto k^2$ is mapped to a linear function of $V_\mathrm{sd}$ in the $V_\mathrm{g}$--$V_\mathrm{sd}$ plane.
Using this relationship, we consider the SBE lines in transconductance $dG/dV_\mathrm{g}$ measurements, in particlular, in the presence of the Zeeman spin-splitting case and the spin--orbit interaction (SOI) spin-splitting case.

\subsection{The wavenumber in actual QPC cases}

\begin{figure} 
\includegraphics[width=0.85\linewidth]{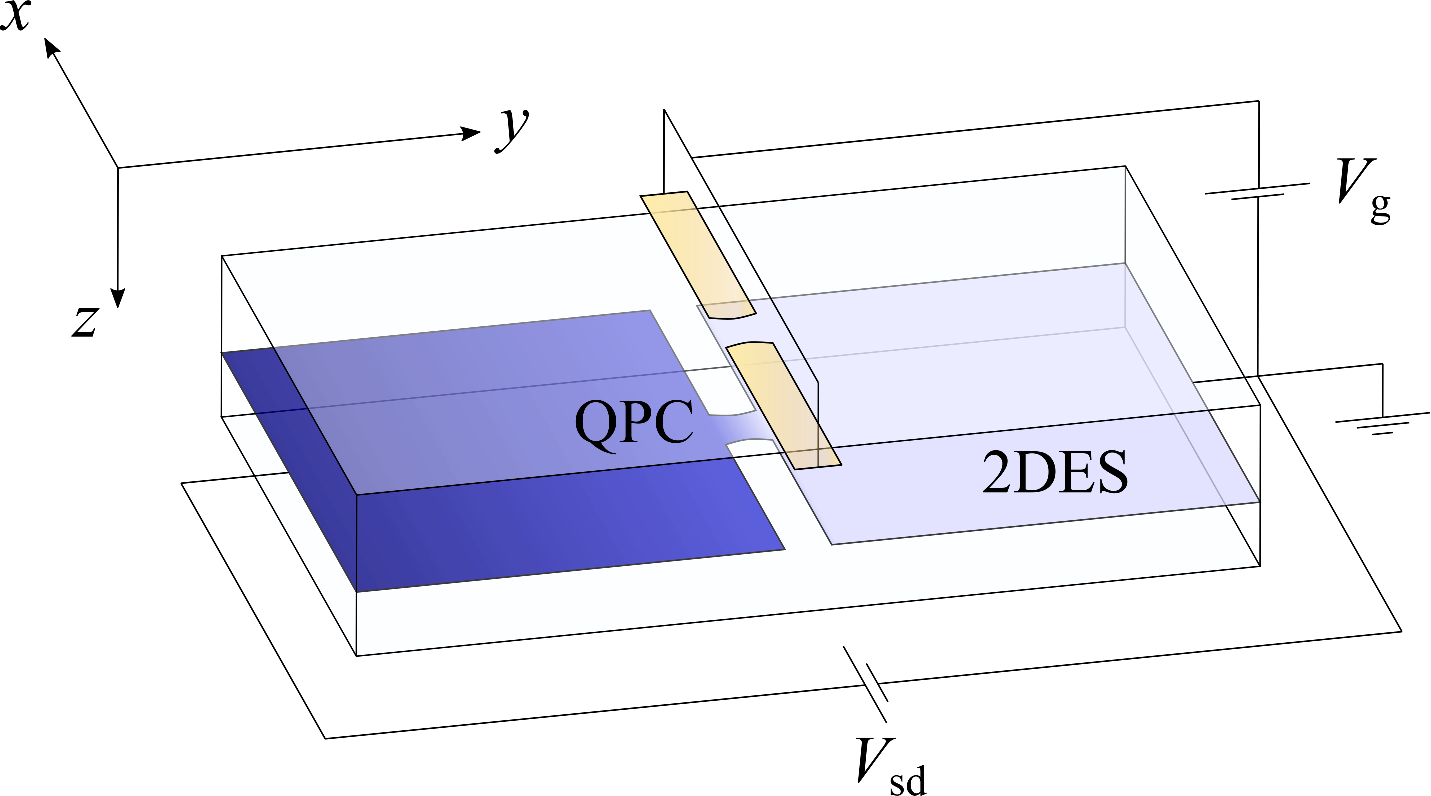}
\caption{\label{fig_QPC_2DEG}(Color online)\,Schematic representation of a QPC configured in a 2DES. }
\end{figure}

As shown in Eq.\,(\ref{wavenumber}), the wavenumber of the accelerated electron depends on $\sqrt{V_\mathrm{sd}}$.
In actual QPC cases, depending on the sign of $V_\mathrm{sd}$, the wavenumber $k_y=p_y/\hbar$ at $y=\gamma L \  (0 \leq \gamma \leq 1)$ becomes
$
k_y = \mathrm{sgn}(V_\mathrm{sd})\sqrt{2\gamma m^\ast |eV_\mathrm{sd}|}/\hbar.       
$
This relationship yields a significant difference when the non-equilibrium bias effect is considered from the model based on the Landauer formula, which assumes that electrons with wavenumbers between $k_\mathrm{d}$ and $k_\mathrm{s}$ transmit a QPC\,\cite{Glazman_1989,Kouwenhoven,Patel_1990}.

Previously, we considered the $y$ direction as the direction of the 1D conduction and the $x$ direction as the lateral direction to the 1D conduction. 
To consider the actual QPC in a 2DES, we take $x$ and $y$ as in-plane components of the 2DES, and the $z$ direction is perpendicular to the 2DES (see Figure\,\ref{fig_QPC_2DEG}).
Then, as discussed earlier, we assumed a quadratic confinement potential $U(x) = m^\ast \omega_x^2x^2/2+U_0$ in the $x$ direction.
Since the narrowest confinement is expected to occur at the center of the QPC region, $\omega_x$ and $U_0$ have the $y$ dependence\cite{Buttiker,Lesovik,Kawabata}, and the quantized energy gap, $\hbar \omega_x$, becomes largest at the center of the QPC ($\omega_x$ becomes largest at $\gamma =0.5$). 
Therefore, we need to consider the wavenumber at the center point.
In fact, as will be shown later, if we consider that the $y$-directional kinetic energy at the center point is equal to the subband spacing, we can explain the relationship between the subband spacing and $V_\mathrm{sd}$ by the vertex of a diamond corresponding to the subband spacing\cite{Kristensen}.
Therefore, in the following we will consider the $\gamma =0.5$ case, 
\begin{equation}
k_y = \mathrm{sgn}(V_\mathrm{sd})\frac{\sqrt{m^\ast |eV_\mathrm{sd}|}}{\hbar}. 
\label{k-V_relation}
\end{equation}

\subsection{Zeeman splitting case}

\begin{figure*}
\includegraphics[width=1\linewidth]{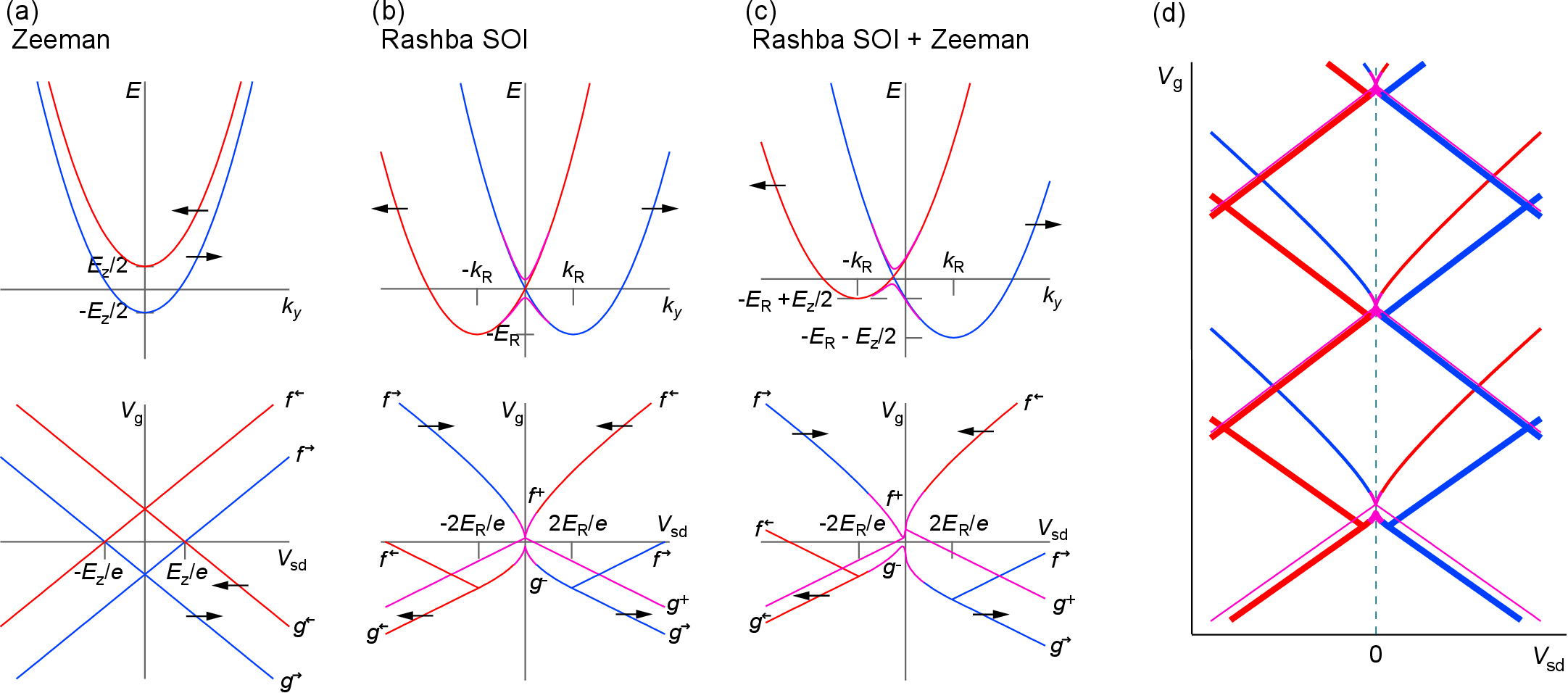}	
\caption{\label{Fig_ZeemanRashba}(Color online)\,Relationship between the dispersion relation and SBE lines in $dG/dV_\mathrm{g}$ $V_\mathrm{g}$--$V_\mathrm{sd}$ plane for (a) the Zeeman splitting, (b) Rashba SOI splitting, and (c) the Rashba SOI plus Zeeman splitting cases. (d) Diamond structures consisting of Rashba SOI splitting cases, as observed in a typical result in Ref.\cite{Kristensen}. }
\end{figure*}

Subsequently, we take the electron spin into account. 
In the presence of an $x$-directional magnetic field $B_x$, the Hamiltonian with the energy gap of Zeeman splitting can be described as follows:
\begin{equation}
\hat{H} =  \frac{\hbar^2\hat{k_y}^2}{2m^\ast} - \frac{1}{2}\sigma_x g^\ast\mu_\mathrm{B}B_x,
\end{equation}
where $\sigma_x$ is the Pauli matrix, $g^\ast$ denotes the Land\'{e}'s $g$-factor (effective $g$-factor), and $\mu_\mathrm{B}$ denotes the Bohr magneton.
Then, if we put $g^\ast\mu_\mathrm{B}B_x = E_\mathrm{Z}$, we have the eigenenergy function as
\begin{equation}
E^\leftrightarrows (k_y) = \frac{\hbar^2k_y^2}{2m^\ast} \pm \frac{E_\mathrm{Z}}{2}.
\end{equation}
By using the relationship in Eq.\,(\ref{k-V_relation}), we can convert the above equation to a function of $V_\mathrm{sd}$ as follows:
\begin{equation}
f^\leftrightarrows (V_\mathrm{sd}) =  \frac{|eV_\mathrm{sd}|}{2} \pm \frac{E_\mathrm{Z}}{2}.
\end{equation}
As these equations explicitely show, the usual quadratic dispersion relation ($\propto k_y^2$) is converted into a linear function of the non-equilibrium bias ($\propto V_\mathrm{sd}$).
If we neglect the ``lever arm" effect and assume that the Fermi energy is proportional to $V_\mathrm{g}$, then the dispersion relation can be mapped as $dG/dV_\mathrm{g}$ peak lines (SBE lines) on the $V_\mathrm{g}$--$V_\mathrm{sd}$ plane as a linear function of $V_\mathrm{sd}$.
In addition, if the total energy of the electrons is equal to the energy gap, they can jump to the upper subband at the lowest possible energy point, which is usually $k_y=0$ for parabolic dispersion. This energy gap decreases with increasing the kinetic energy ($=\frac{\hbar^2k_y^2}{2m^\ast} =\frac{|eV_\mathrm{sd}|}{2}$), thus by subtracting the kinetic energy from the gap, it appears in the $V_\mathrm{g}$--$V_\mathrm{sd}$ plane as
\begin{equation}
g^\leftrightarrows (V_\mathrm{sd}) = -\frac{|eV_\mathrm{sd}|}{2} \pm \frac{E_\mathrm{Z}}{2}.
\end{equation}
The dispersion relation and the corresponding $dG/dV_\mathrm{g}$ lines in the $V_\mathrm{g}$--$V_\mathrm{sd}$ plane for the Zeeman splitting case are displayed in Fig.\,\ref{Fig_ZeemanRashba} (a).
This also applies to the case of lateral quantization with the gap of $\hbar \omega_x$.
Consequently, we can explain the diamond structure of the SBE lines in the $V_\mathrm{sd}$--$V_\mathrm{g}$ plane observed in several QPC experiments\cite{Thomas_PRB1998,Kristensen,Cronenwett,DiCarlo,dePicciotto,Roessler,Fischer,Martin_PRB,Chen_NanoLett,Crook,Yan_PRB,bilayerQPC}.
The crossing points are at $V_\mathrm{sd} = \pm E_\mathrm{Z}/e$, thus, as in Ref.\,\cite{Kristensen}, we can deduce the magnitude of the Zeeman gap or the lateral quantization gap ($\hbar\omega_x$) from the diamond structure.

\subsection{SOI splitting case}

\begin{figure}
\includegraphics[width=1\linewidth]{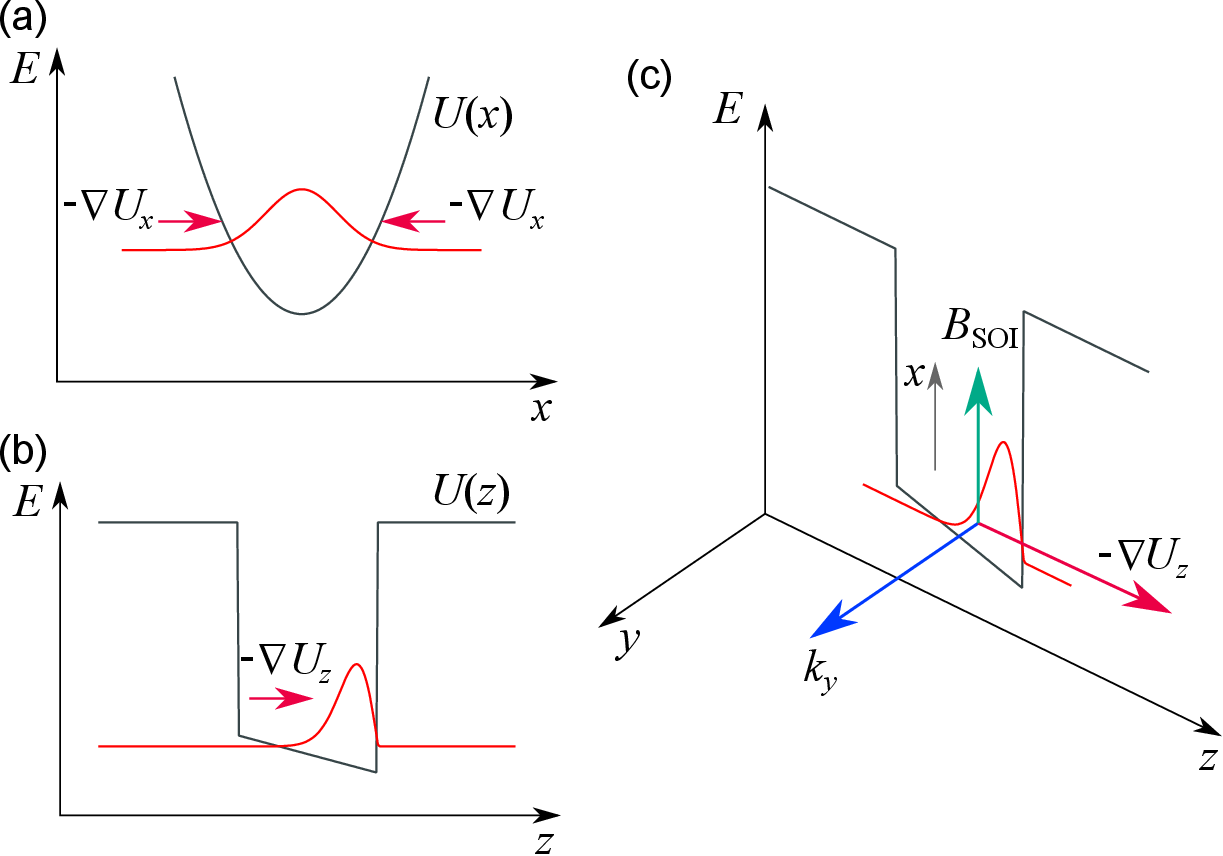}
\caption{\label{fig_gradV_x_ky}(Color online)\,(a) Schematic representation of the $x$-directional potential function $U(x)=m^\ast \omega^2x^2/2$ and its gradient in QPC. (b) Schematic representation of the $z$-directional potential function $U(z)$ and its gradient in the case of a quantum well. (c) The SOI field $B_\mathrm{SOI}$ is from $[\nabla U \times \bm{k}]_x=-\frac{\partial U}{\partial z}k_y$.}
\end{figure}

Now, let us consider a case in which a dispersion relation is modified by a $k$-directional splitting, such as SOI splitting dispersion relations, as shown in Fig.\,\ref{Fig_ZeemanRashba} (b). 
SOI splitting can occur in the presence of asymmetric potential modulations perpendicular to the wave vector. 
In the case of 2DES in quantum wells and at interfaces of GaAs/AlGaAs heterostructures, electrons experience a $z$-directional potential gradient, as shown in e.g. \cite{bilayerQPC} (see also Fig.\,\ref{fig_gradV_x_ky}).
As discussed in \cite{bilayerQPC}, only the $z$-directional spatial symmetry is explicitly broken;  regarding the $x$-direction, the potential gradients on either side of a parabolic confinement potential cancel each other out.
Otherwise, the $x$-directional symmetry can be broken if we apply different voltages to each split gate\cite{Debray}; however, we do not consider this case in this study.
For a system with a $z$-directional potential gradient and a $y$ directional current flow, a Rashba SOI in the $x$-direction is expected due to the vector product, $ [0,0, \frac{\partial U(z)}{\partial z}] \times [0, k_y, 0] = [-\frac{\partial U(z)}{\partial z} k_y, 0, 0] $. Thus, the Rashba SOI Hamiltonian can be expressed as follows:
\begin{align}
\hat{H} &= \frac{\hbar^2\hat{k_y}^2}{2m^\ast} -\frac{\hbar^2}{4{m^\ast}^2c^2}\sigma_x \frac{\partial U(z)}{\partial z}\hat{k_y}  \nonumber \\ &= \frac{\hbar^2\hat{k_y}^2}{2m^\ast} - \sigma_x \alpha_\mathrm{R}\hat{k_y},      \label{Hamiltonian}
\end{align}
where $U(z)$ denotes the potential of the quantum well and $\alpha_\mathrm{R}$ is the so-called Rashba parameter. 
The dispersion relation is derived from the eigenvalues as follows:
\begin{equation}
E^\leftrightarrows(k_y) = \frac{\hbar^2k_y^2}{2m^\ast} \pm \alpha_\mathrm{R}k_y.  \label{dispersion}
\end{equation}
In this dispersion relation, the energy has a minimum value of $ - \hbar^2k_\mathrm{R}^2/(2m^\ast) = -E_\mathrm{R}$ at $k_y= \mp \frac{m^\ast \alpha_\mathrm{R}}{\hbar^2} = \mp k_\mathrm{R}$.
Furthermore, these branches repel each other at the crossing point and open a small gap due to the perturbation effect\,\cite{Goulko} as 
\begin{align}
E^\pm (k_y) &= \frac{1}{2} \left[ E^\leftarrow + E^\rightarrow \pm \sqrt{(E^\leftarrow -  E^\rightarrow)^2 + 4|V_{12}|^2} \right] \nonumber \\
		&= \frac{\hbar^2k_y^2}{2m^\ast} \pm \sqrt{\alpha_\mathrm{R}^2k_y^2 +V_{12}^2}, \label{perturbation}
\end{align}
where $V_{12}$ is the gap between the upper and lower branches.
Consequently, the SBE lines in the $V_\mathrm{g}$--$V_\mathrm{sd}$ plane becomes more complicated, because the minima do not occur at $k_y=0$, that is, $V_\mathrm{sd}=0$.

By converting $k_y$ to $V_\mathrm{sd}$ using Eq.\,(\ref{k-V_relation}) and using $\alpha_\mathrm{R}k_y =\sqrt{2E_\mathrm{R}|eV_\mathrm{sd}|}$, the SBE lines belonging to the upper branches $E^+$, $E^\leftarrow (k_y>0)$, and $E^\rightarrow (k_y<0)$ become as follows:
\begin{gather}
f^+ = 	\frac{|eV_\mathrm{sd}|}{2} + \sqrt{2 E_\mathrm{R} |eV_\mathrm{sd}| + |V_{12}|^2}   \quad   V_\mathrm{sd} \sim 0 \\
f^\leftrightarrows	= \frac{|eV_\mathrm{sd}|}{2} + \sqrt{2 E_\mathrm{R}|eV_\mathrm{sd}|}    \quad   \text{otherwise.}
	 \label{SBERashba_Upper_Rise} \\
g^+ = - \frac{|eV_\mathrm{sd}|}{2} + V_{12} ,  \label{SBERashba_Upper_JumpIn}
\end{gather}
Regarding the lower dispersion branches $E^-$, $E^\leftarrow (k_y <0)$, and $E^\rightarrow (k_y >0)$, it should be noted that these branches have two minima; thus, we need to separate the case at the minima. 
If we note that the minimum point of these branches varies as a function of $V_\mathrm{sd}$ for $|V_\mathrm{sd}| \leq 2E_\mathrm{R}/e$, the kinetic energy term cancels out in this region and the energy gap becomes
\begin{gather}
g^- = - \sqrt{2 E_\mathrm{R} |eV_\mathrm{sd}| + |V_{12}|^2} \quad  V_\mathrm{sd} \sim 0 \\
g^\leftrightarrows = 
    \begin{cases}
	- \sqrt{2 E_\mathrm{R}|eV_\mathrm{sd}|}   \quad  &  0 \ll |V_\mathrm{sd}| < \frac{2E_\mathrm{R}}{e}  \\
	- E_\mathrm{R} - \frac{|eV_\mathrm{sd}|}{2} & |V_\mathrm{sd}| \ge \frac{2E_\mathrm{R}}{e}. 
	\end{cases}
\end{gather}
From the two minima, we can consider the following lines corresponding to $E^\leftrightarrows$:
\begin{equation}
f^\leftrightarrows = - 3E_\mathrm{R} + \frac{|eV_\mathrm{sd}|}{2}  \ \quad  |V_\mathrm{sd}| \ge \frac{2E_\mathrm{R}}{e}.
\end{equation}
Consequently, we obtained a combination of SBE lines as shown in Figure \ref{Fig_ZeemanRashba} (b).

This model attributes the controversial small plateau observed at approximately $0.7\cdot 2e^2/h$ to the perturbation effect: the perturbation effect opens a small gap at the crossing point ($V_\mathrm{sd} =0$) of the SOI dispersion relation, which causes a small plateau-like conductance change.  
Furthermore, the peculiar SBE lines in the $V_\mathrm{g}$--$V_\mathrm{sd}$ plane, observed in many experiments such as\cite{Thomas_PRB1998,Kristensen,Cronenwett,DiCarlo,dePicciotto,Martin_PRB,Chen_NanoLett,Crook,Yan_PRB}, have been comprehensively reproduced.
The important features that agree with the experimental results are as follows:
First, the SBE lines of the first diamond do not converge to $V_\mathrm{sd}=0$\,V, because the minima of the Rashba SOI dispersion deviate from $V_\mathrm{sd} = 0$. Second, the two increasing lines from the 0.7 plateau increase rapidly at first because they increase as shown in Eq.\,(\ref{SBERashba_Upper_Rise}). 
Consequently, typical SBE line structures in the $V_\mathrm{g}$--$V_\mathrm{sd}$ plane are reproduced as a result of multiple SOI splitting subbands that are separated by $\hbar\omega_x$ as shown in Fig.\,\ref{Fig_ZeemanRashba} (d).
Subsequently, we can attribute three converging points at $V_\mathrm{sd}=0, \pm 2E_\mathrm{R}/e$ ($k_y =0, \pm k_\mathrm{R}$), which originate from the intersection and two minima of the SOI dispersion relation, to the three peaks in $G$ observed in many experiments such as Refs.\cite{Kristensen,Cronenwett,Chen_PRB,Sarkozy,bilayerQPC}.
Among them, the center peak is known as the zero bias anomaly\cite{Cronenwett,Bauer}. 
The other two peaks appear before the center peak with much stronger intensity, because these two peaks belong to lower energy points than the center peak.
These points also appear in $dG/dV_\mathrm{g}$ peaks, because the conductance changes rapidly at these points.
As an important consequence, this model suggests that the spin can be easily separated by changing the direction of the current.

In this $k$-directional splitting model, the crossing point at $k_y=0\ (V_\mathrm{sd}=0)$ is protected by the time-reversal symmetry (Kramers point), therefore it is difficult to lift the spin degeneracy at this point.
To lift this degeneracy at $V_\mathrm{sd}=0$, applying a magnetic field is nonetheless a feasible method.
Here, we consider a situation in the presence of a magnetic field parallel to the Rashba SOI field  ($B_x$ in this study). 
The Zeeman term is added to the Hamiltonian of Eq.\,(\ref{Hamiltonian}) as 
\begin{equation}
\hat{H} 
= \frac{\hbar^2\hat{k_y}^2}{2m^\ast} - \sigma_x\alpha_\mathrm{R}\hat{k_y} - \frac{1}{2}\sigma_x g\mu_\mathrm{B}B_x.     \label{Hamiltonian_withZeeman}
\end{equation}
The dispersion relation from this Hamiltonian becomes as follows:
\begin{equation}
E^\leftrightarrows(k_y) = \frac{\hbar^2 k_y^2}{2m^\ast} \pm \alpha_\mathrm{R}k_y \pm \frac{E_\mathrm{z}}{2}.     \label{RashbaZeemanDispersion_leftright}
\end{equation}
Likewise Eq. (\ref{perturbation}), the perturbation effect splits the dispersion in the vicinity of the crossing point as
\begin{equation}
E^\pm (k_y) = \frac{\hbar^2 k_y^2}{2m^\ast} \pm \sqrt{\alpha_\mathrm{R}^2k_y^2 + \alpha_\mathrm{R}k_y E_\mathrm{z} + E_\mathrm{z}^2/4 + V_{12}^2}.   \label{RashbaZeemanDispersion_perturb}
\end{equation}
As a result of horizontal and longitudinal splitting, the subband gap opens differently between the positive and negative sides of $V_\mathrm{sd}$.
This deviates the crossing point to the positive or negative side depending on the direction of the applied magnetic field ($B_x$) and the SOI field. 
Consequently, the gap opens at $V_\mathrm{sd} = 0$. 
However, this requires a strong magnetic field.
The magnetic field varies the crossing point to the negative or positive side with respect to the applied direction, as
\begin{equation}
k_y^\mathrm{c} = - \frac{E_\mathrm{Z}}{2\alpha_\mathrm{R}},         \label{ky_c}
\end{equation}
thus, we have a deviated crossing point at 
\begin{equation}
V_\mathrm{sd}^\mathrm{c} = - \mathrm{sgn}(E_\mathrm{Z})\frac{E_\mathrm{Z}^2}{8eE_\mathrm{R}}.           \label{Vsd_c}
\end{equation}
It should be noted that $E_\mathrm{Z} = g^\ast\mu_\mathrm{B}B_x$ and $B_x$ can be either negative or positive. 
In ref.\cite{Thomas}, this was shown that a small $0.7 \cdot 2e^2/h$ conductance structure develops to two larger plateaus of $0.5 \cdot 2e^2/h$ and $1 \cdot 2e^2/h$ at approximately 9\,T.  
Furthermore, the direct observation of this crossing point shift was probably reported by Ref.\,\cite{bilayerQPC}, in that the three maxima points obeyed the $B_x$ dependence of this model.
The SBE lines corresponding to Eqs. (\ref{RashbaZeemanDispersion_leftright}) and (\ref{RashbaZeemanDispersion_perturb}) are shown in the bottom part of Fig\,\ref{Fig_ZeemanRashba}(c) 
(For the detailed derivation, see Appendix \ref{RashbaPlusZeeman_Eq}).

Furthermore, it is noteworthy that many experimental results show a small asymmetry with respect to the inversion of $V_\mathrm{sd}$ at zero magnetic field\cite{Kristensen,Cronenwett,bilayerQPC}.
For example, Kristensen {\it et al}.\cite{Kristensen} reported higher conductance for the negative side of $V_\mathrm{sd}$ than for the positive side, although they explained this asymmetry in terms of a self-gating effect. 
This asymmetry was also reported in shot noise measurements\cite{Roche,bilayerQPC}, in which an asymmetric noise increase and a Fano factor increase were observed with respect to $V_\mathrm{sd}$.
This can be attributed to spin splitting due to the effective field created by the Rashba SOI ($\propto g^\ast\mu_\mathrm{B}B_\mathrm{SOI}$).
Otherwise, we infer that this observation is a signature of spontaneous spin polarization due to the electron-electron interaction, which has been discussed in many theoretical studies in the 1D region\cite{WangBerggrenR,WangBerggren,Reilly,DaulNoack,Yang1D,Aryanpour}.
However, the interaction was shown to be negligibly small in the 1D region\,\cite{Tarucha_SSC}. Thus, the interaction in the 2DES region should be considered in this model, because, as mentioned in Sec.\,\ref{QPCsIn2DES}, a homogeneous field locally breaks the charge neutrality condition due to charge accumulation, which leads to an enhancement of the electron-electron interaction in this region.
The importance of the interaction between the 2D electrons has been discussed in several studies\cite{MaslovStone,Ponomarenko,SafiSchulz}.
Discussion that includes the electron-electron interaction using, for example, the mean-field approximation is necessary.
There is also an urgent need for further theoretical discussion on quantitative analyses of the Rashba parameter and conductance values considering temperature dependence.

\section{Concluding remarks}

In summary, we have derived the quantized conductance from the accelerated electron model. 
In this model, the fundamental property of the quantized conductance is that a finite amount of time is required to carry electric charges by electrons in a uniformly accelerated motion. 
As a result, the quantized conductance is an intrinsic property of a 1D ballistic conductor.
We also discuss the possibility that the finite conductance does not require a finite amount of dissipation in the form of Joule heat, because the work done by the electric field is used to increase the kinetic energy of the electrons. 
Thus, it is an energy-loss-free system.
As a comparison with a realistic model, we discuss the similarity between a QPC and an hourglass. 
We have shown that the nodal spacings in the coherent electron flows emitted by QPCs are consistent with the de-Broglie wavelength derived from this model.
Furthermore, using the relationship between the wavenumber of electrons and the applied DC bias $V_\mathrm{sd}$ across a QPC, 
we were able to reproduce the peculier SBE lines observed in association with the 0.7 anomaly.   
Consequently, the 0.7 anomaly is understood as a perturbation gap between the Rashba SOI splitting dispersion curves at $k=0$. 

As we have seen, the discussion ignores the many-body electron-electron interaction, which is often crucial to understand phenomena of the condensed matter physics.
However, we believe that it is important to explain the physics in the framework of the single-electron model first, as it always a key to the clear understanding. 
We have raised a possible case of the many-body interaction briefly, however, the interaction  takes place in the 2DES in this case. 
Simple considerations suggest that the Coulomb interaction between the electrons prevents them from starting simultaneously, however, although this is a slight repetition of the literature\cite{NumOfElectrons}, it scarcely affects the accelerated motion in a homogeneous field.
Spin interaction is also negligible because the eigenenergy of electrons in motion is different, therefore exempt from the Pauli's exclusion principle.
Further consideration is necessary, however, thorough discussion and formulation are beyond the scope of this study, because the aim of this study is to establish a single-electron model.

We believe that the derivation and interpretation proposed here contributes to the further understanding of the quantized conductance and the physics of the QPC, as well as to the physics of non-equilibrium conductions.
Further quantitative analyses, such as the Rashba parameter and the 0.7 plateau value, are required. 
The results of this study are profitable for spintronic applications proposed by Datta-Das\cite{DattaDas}, such as spin filtering and manipulation for quantum engineering, in well-developed GaAs/AlGaAs compound semiconductor platforms.
This model proposes more versatality to experiments, offerring a nano-meter-size electron accelerator, wavelength modulator, or coherent spin flow generator.
An important task to verify should be the spin splitting effect caused by the SOI.

\section*{Acknowledgements}
The author would like to thank A. Ueda for carefully reading and commenting on this manuscript.

\begin{appendix}

\section{Quantized conductance in a linear band}
\label{LinearBand}

As a further example, we show that the method for deriving the quantized conductance for a massive dispersion case can also be adapted to a massless linear band case.
When the band is linear, the relationship between the energy $E$ and the momentum $p$ is $E = cp = c\hbar k$ (where $c$ is the constant group velocity in the band), and the momentum increases as $\dot{p} = F$.
When $E = eV_\mathrm{sd}$, we obtain $k_L=eV_\mathrm{sd}/(c\hbar)$. 
Consequently, as $\tau=L/c$, the current becomes
\begin{equation}
J=\frac{eN}{\tau}=\frac{e}{L/c}\int_0^{k_L} 2L \frac{dk}{2\pi} = \frac{e}{L/c}\frac{2LeV_\mathrm{sd}}{2\pi c\hbar}=\frac{2e^2}{h}V_\mathrm{sd}.
\end{equation}
Therefore, the quantized conductance is also derived for a linear band case.
Note that in this case the relation between $k$ and $V_\mathrm{sd}$ is $k \propto V_\mathrm{sd}$.

It is known that graphene strips exhibit a quantized conductance\cite{Tombros2011}.
This result is another example of the validity of this theory, which is based on uniformly accelerated motion, or more precisely, we should call it uniformly increased momentum motion.
However, as already discussed, the electrons must satisfy the quantization condition, which causes the electronic band dispersion to change hyperbolically, because a vertical section of a cone becomes a hyperbola.
Since the hyperbola can be approximated by a quadratic curve in the low energy region, it eventually reduces to a case of massive dispersion. 

\section{Dispersion relation and SBE lines for the Rashba SOI + Zeeman case}
\label{RashbaPlusZeeman_Eq}

Subsequent to Eqs.\,(\ref{RashbaZeemanDispersion_leftright}) and (\ref{RashbaZeemanDispersion_perturb}), we convert $k_y$ to $V_{\rm sd}$ according to $k_y=\mathrm{sgn}(V_{\rm sd})\sqrt{m^\ast|eV_{\rm sd}|}/\hbar$. 
As shown in Eqs.\,(\ref{ky_c}) and (\ref{Vsd_c}), when $E_{\rm z} >0$, the crossing point shifts to the negative side, and if $E_{\rm z} < 0$, it shifts to the positive side. 
Thus the corresponding SBE lines as a function of $V_{\rm sd}$ belonging to the upper branches $E^+$, $E^\leftarrow (k_y >0)$, and $E^\rightarrow (k_y < 0)$ become as follows:
\begin{widetext}
\begin{gather}
f^+ = 
\frac{|eV_{\rm sd}|}{2} +
\sqrt{2E_{\rm R}|eV_{\rm sd}| + \mathrm{sgn}(V_{\rm sd})E_{\rm z} \sqrt{2E_{\rm R} |eV_{\rm sd}|} + \frac{E_{\rm z}^2}{4} + V_{12}^2}  \quad   V_{\rm sd} \sim V_{\rm sd}^{\rm c}, \\
f^\leftarrow = \frac{|eV_{\rm sd}|}{2} + \sqrt{2E_{\rm R} |eV_{\rm sd}|} +\frac{E_{\rm z}}{2}   \quad  V_{\rm sd} \gg V_{\rm sd}^{\rm c}, \\
f^\rightarrow = \frac{|eV_{\rm sd}|}{2} + \sqrt{2E_{\rm R} |eV_{\rm sd}|} - \frac{E_{\rm z}}{2}  \quad   V_{\rm sd} \ll V_{\rm sd}^{\rm c}, 
\end{gather}
where $E_{\rm R}=m^\ast \alpha_{\rm R}^2/(2\hbar^2)$. 
The cases for the jumping-in electrons for $E_{\rm z} >0$ (for $E_{\rm z} <0$) are:
\begin{equation}
g^+ =
\begin{cases} 
		\sqrt{E_{\rm z}^2/4 +V_{12}^2} -\frac{|eV_{\rm sd}|}{2}       &V_{\rm sd} \ge 0 \quad(V_{\rm sd} \le 0) \\
		\sqrt{ E_{\rm R}|eV_{\rm sd}| + \mathrm{sgn}(V_{\rm sd})E_{\rm z}\sqrt{E_{\rm R}|eV_{\rm sd}|} + \frac{E_{\rm z}^2}{4} + V_{12}^2} 		 &V_{\rm sd}^{\rm c} \le V_{\rm sd} < 0 \quad(0 < V_{\rm sd} \le V_{\rm sd}^{\rm c} )  \\
		- \frac{|eV_{\rm sd}|}{2} + V_{12} + \frac{E_{\rm z}^2}{8E_{\rm R}} & V_{\rm sd} < V_{\rm sd}^{\rm c} \quad(V_{\rm sd} > V_{\rm sd}^{\rm c}).
\end{cases}
\end{equation}
For the lower branch $E^-$, $E^\leftarrow (k_y<0)$, and $E^\rightarrow (k_y >0)$, the jumping-in SBE lines become
\begin{equation}
g^\leftarrow =
\begin{cases}
		- \frac{|eV_{\rm sd}|}{2} -E_{\rm R} + \frac{E_{\rm z}}{2}	 & \text{for}\ V_{\rm sd} \le -\frac{2E_{\rm R}}{e} \\
		- \sqrt{2E_{\rm R}|eV_{\rm sd}|} + \frac{E_{\rm z}}{2} 	 & \text{for}\ -\frac{2E_{\rm R}}{e} < V_{\rm sd} \ll V_{\rm sd}^{\rm c}
\end{cases}
\end{equation}
\begin{equation}
g^-=	- \sqrt{2E_{\rm R}|eV_{\rm sd}| + \mathrm{sgn}(V_{\rm sd})E_{\rm z}\sqrt{2E_{\rm R}|eV_{\rm sd}|} +\frac{E_{\rm z}^2}{4} + V_{12}^2}  \quad  V_{\rm sd} \sim V_{\rm sd}^{\rm c} 
\end{equation}
\end{widetext}
\begin{equation}
g^\rightarrow =
\begin{cases}
		-\sqrt{2E_{\rm R}|eV_{\rm sd}|} -\frac{E_{\rm z}}{2} 	& V_{\rm sd}^{\rm c} \ll V_{\rm sd} < \frac{2E_{\rm R}}{e} \\
		- \frac{|eV_{\rm sd}|}{2} -E_{\rm R} - \frac{E_{\rm z}}{2}   & \frac{2E_{\rm R}}{e}\le V_{\rm sd}. 
\end{cases}
\end{equation}
In addition, we have increasing lines for $|V_{\rm sd}| \ge E_{\rm R}/e$, such that 
\begin{gather}
f^\leftarrow  = 
	\frac{|eV_{\rm sd}|}{2} - 3E_{\rm R} + \frac{E_{\rm z}}{2} \quad V_{\rm sd} \le -\frac{2E_{\rm R}}{e} \\ 
f^\rightarrow  =
	\frac{|eV_{\rm sd}|}{2} - 3E_{\rm R} - \frac{E_{\rm z}}{2} \quad V_{\rm sd} \ge \frac{2E_{\rm R}}{e}. 
\end{gather}
The SBE lines of these equations are shown in Fig.\,\ref{Fig_ZeemanRashba} (c).

\end{appendix}

\bibliographystyle{unsrt.bst}

\begin{thebibliography}{85}
\bibitem [{not()}]{note_quantizedwavenumber}%
  \BibitemOpen
  \href@noop {} {}\bibinfo {note} {The wavenumber is quantized if we take the
  infinitely deep quantum well potential as the confinement potential, however,
  we remain ourselves to the harmonic oscillator to discuss more realistic
  model.}\BibitemShut {Stop}%
\end{thebibliography}

%

\end{document}